\documentclass[journal]{IEEEtran}
\usepackage[utf8]{inputenc}
\usepackage{amsmath}
\usepackage{amsfonts}
\usepackage{amssymb}
\usepackage{graphicx}
\usepackage{cite}
\usepackage{url}
\usepackage{booktabs}
\usepackage{array}
\usepackage{multirow}
\usepackage{tikz}
\usepackage{pgfplots}
\usepackage{setspace}
\usepackage{threeparttable}
\pgfplotsset{compat=1.18}

\usepackage{fancyhdr}
\fancypagestyle{plain}{%
  \fancyhf{}

  \fancyfoot[C]{\thepage}
}
\title{Evaluating Device-First Continuum AI (DFC-AI) for Autonomous Operations in the Energy Sector}
\author{\IEEEauthorblockN{Siavash M. Alamouti\IEEEauthorrefmark{1}, Fay Arjomandi\IEEEauthorrefmark{1}, Michel Burger\IEEEauthorrefmark{1}, and Bashar Altakrouri\IEEEauthorrefmark{2}}
\IEEEauthorblockA{\IEEEauthorrefmark{1}mimik Technology, USA}
\IEEEauthorblockA{\IEEEauthorrefmark{2}Saudi Aramco, Saudi Arabia}}

\begin{document}
\maketitle

\begin{abstract}
Industrial automation in the energy sector requires AI systems that can operate autonomously regardless of network availability, a requirement that cloud-centric architectures cannot meet. This paper evaluates the application of Device-First Continuum AI (DFC-AI) to critical energy sector operations. DFC-AI, a specialized architecture within the Hybrid Edge Cloud paradigm, implements intelligent agents using a microservices architecture that originates at end devices and extends across the computational continuum. Through comprehensive simulations of energy sector scenarios including drone inspections, sensor networks, and worker safety systems, we demonstrate that DFC-AI maintains full operational capability during network outages while cloud and gateway-based systems experience complete or partial failure. Our analysis reveals that DFC-AI's zero-configuration GPU discovery and heterogeneous device clustering are particularly well-suited for energy sector deployments, where specialized nodes can handle intensive AI workloads for entire fleets of inspection drones or sensor networks. The evaluation shows that DFC-AI achieves significant latency reduction and energy savings compared to cloud architectures. Additionally, we find that gateway-based edge solutions can paradoxically cost more than cloud solutions for certain energy sector workloads due to infrastructure overhead, while DFC-AI can consistently provide cost savings by leveraging enterprise-owned devices. These findings, validated through rigorous statistical analysis, establish that DFC-AI addresses the unique challenges of energy sector operations, ensuring intelligent agents remain available and functional in remote oil fields, offshore platforms, and other challenging environments characteristic of the industry.
\end{abstract}


\section{Introduction: The Imperative for Always-Available Intelligence}

The promise of AI in industrial automation rests on a fundamental requirement: intelligent agents must be available anytime, anywhere, regardless of network conditions. In the energy sector, where operations span remote oil fields, offshore platforms, and hazardous environments, this requirement is not negotiable. It is essential for both operational continuity and human safety. Yet traditional cloud-centric AI architectures fail catastrophically when network connectivity is lost or degraded, rendering intelligent systems inoperative precisely when they may be needed most~\cite{sterling2022ai,motion2022edge}.

Consider the operational landscape within the energy sector. Autonomous robots inspecting pipelines, drones monitoring offshore platforms, smart sensors deployed across oilfields, and control systems within refineries all generate large amounts of contextual data directly at the point of operation. For these devices to make accurate and timely decisions which may be crucial for safety, efficiency, and preventing costly downtime, they require immediate access to this local context. Transferring compressed video streams and sensor data, ranging from hundreds of megabytes to several gigabytes per device per day~\cite{zhou2024survey,chowdhury2024statistical,riesner2022analysis}, to a centralized cloud for processing before a decision can be made introduces unacceptable latency, rendering real-time control and intervention impossible in many critical situations. Furthermore, the nature of many industrial AI workloads follows a Pareto distribution, with approximately 80\% of decisions being localized and context-dependent~\cite{barroso2013datacenter,xu2017cloud,reiss2011google}, making them ideally suited for processing at the device level.

The challenge extends beyond current operations. As the industry transitions from millions of human-operated applications to billions of autonomous AI agents, the architectural limitations of centralized approaches become insurmountable. Simple mathematics illustrates this scalability crisis: one billion autonomous agents each requiring modest 1MB/second communication with cloud services would demand 1,000 TB/second of aggregate network bandwidth, exceeding the capacity of entire continental internet infrastructure. This fundamental barrier demands a complete architectural rethink.

To address these demands, a paradigm shift towards Hybrid Edge Cloud~\cite{alamouti2022hybrid,alamouti2024building} architectures is gaining momentum~\cite{foley2022hybrid}. Within this paradigm, DFC-AI~\cite{alamouti2025dfc} emerges as a specialized approach. DFC-AI is characterized by the implementation of intelligent agents, architected using a microservices framework, that are initiated and primarily reside on devices. These intelligent agents can then extend their operations and collaborate with gateways and cloud servers as needed, forming a cohesive and distributed intelligence continuum. Similar to how the human brain handles immediate reflexes and local tasks at the peripheral level while relying on the central nervous system for complex reasoning and global awareness, DFC-AI prioritizes intelligence at the device level for timely actions while leveraging the broader cloud continuum for more involved processes.

This paper presents comprehensive evidence for DFC-AI's advantages through mathematical modeling, detailed simulations of real-world energy sector scenarios with statistical validation, and verification of theoretical predictions. We demonstrate not only the performance benefits but, more importantly, how DFC-AI ensures that industrial AI systems deliver on their fundamental promise: intelligent automation that is always available when and where it is needed.

\section{Background: DFC-AI Architecture and Energy Sector Requirements}

\subsection{Device-First Continuum AI Overview}

Device-First Continuum AI (DFC-AI)~\cite{alamouti2025dfc} represents a paradigm shift from traditional cloud-centric approaches by prioritizing intelligence at the device level. As described in~\cite{alamouti2025dfc}, DFC-AI implements intelligent agents using a microservices architecture that originates at end devices and extends across the computational continuum. The architecture is built on the Hybrid Edge Cloud paradigm~\cite{alamouti2022hybrid,alamouti2024building} and enables devices to process data locally while maintaining the ability to collaborate with edge gateways and cloud servers when beneficial.

Key characteristics of DFC-AI relevant to energy sector applications include:

\begin{itemize}
\item \textbf{Intelligence at the Source:} AI capabilities embedded directly on devices enable real-time processing without network dependency, critical for remote energy operations.

\item \textbf{Microservices Architecture:} Lightweight, modular agents that activate only required components, optimizing energy consumption in battery-powered devices.

\item \textbf{Zero-Configuration Resource Discovery:} Automatic discovery and utilization of computational resources within the local environment, enabling dynamic clustering of inspection drones or sensor networks.

\item \textbf{Network Resilience:} Full operational capability maintained during network outages, essential for safety-critical systems in isolated locations.

\item \textbf{Heterogeneous Device Support:} Efficient utilization of mixed computational capabilities, where GPU-equipped devices handle intensive workloads for entire fleets.
\end{itemize}

\subsection{Energy Sector Operational Requirements}

The energy sector presents specific challenges that make traditional cloud-centric AI architectures inadequate:

\textbf{1. Remote and Hazardous Locations:} Oil fields, offshore platforms, and pipeline networks often operate in areas with limited or unreliable network connectivity. Systems must function autonomously for extended periods.

\textbf{2. Safety-Critical Operations:} Worker safety monitoring, gas leak detection, and emergency response systems cannot tolerate network-induced delays or failures. Immediate local response is essential.

\textbf{3. Massive Data Generation:} Modern energy operations generate enormous data volumes. A single inspection drone can produce hundreds of megabytes daily from compressed high-resolution imaging and sensor readings~\cite{zhou2024survey}.

\textbf{4. Real-Time Control Requirements:} Process control in refineries, valve adjustments in pipelines, and collision avoidance for autonomous vehicles require sub-second response times.

\textbf{5. Economic Constraints:} Energy companies require cost-effective solutions that leverage existing infrastructure investments rather than requiring continuous cloud service payments.

These requirements dictate an architectural approach that ensures autonomous operation, minimizes latency, reduces data transmission costs, and maintains functionality regardless of network availability.

\section{Architectural Comparison: DFC-AI vs. Gateway-Based Edge AI vs. Centralized Cloud AI}

The selection of an appropriate architectural paradigm is fundamental to the successful implementation of intelligent systems in industrial automation. This section outlines the main differences between three primary architectural approaches: Centralized Cloud AI, Gateway-Based Edge AI, and DFC-AI.

\subsection{Centralized Cloud AI}

\begin{figure}[htbp]
\centering
\includegraphics[width=\columnwidth]{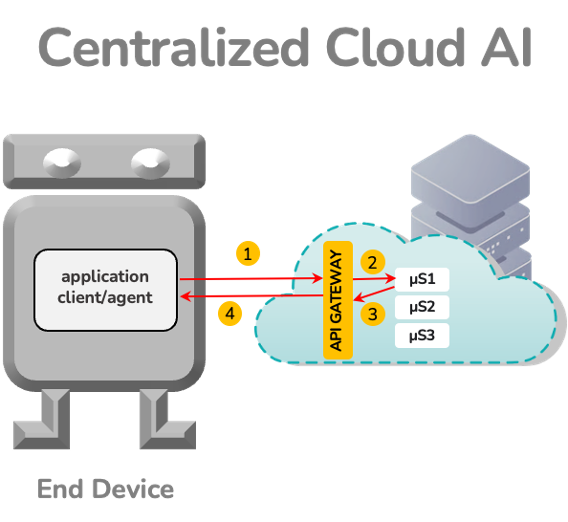}
\caption{High Level Architecture of Centralized Cloud Native AI Solutions.}
\label{fig:centralized_cloud}
\end{figure}

In this architecture shown in Figure~\ref{fig:centralized_cloud}, all data from end devices is transmitted to a central cloud infrastructure for processing and analysis. The end device has no ability to host workloads and only includes a client agent with or without user interface depending on the type of device.

While the cloud offers computational resources and scalability, this approach suffers from several critical limitations. Most significantly, it exhibits complete operational failure during network outages; when internet connectivity is lost, all intelligent functionality ceases, rendering devices non-functional. This creates a single point of failure that is particularly problematic in remote industrial environments where network reliability cannot be guaranteed. Additionally, this approach suffers from high communication latency, making it unsuitable for applications requiring real-time responses. It requires continuous and reliable network connectivity and can lead to substantial bandwidth consumption and cloud hosting costs, especially with large deployments and high data volumes. The context and operational data that must be continuously transferred to the cloud can range from gigabytes to terabytes per day, amplifying bandwidth requirements and associated costs. Data privacy and security can also be a concern as sensitive operational data is transmitted over the network and resides in a centralized environment. End devices in this model typically have very low autonomy, acting primarily as data conduits.

\subsection{Gateway-Based Edge AI}

\begin{figure}[htbp]
\centering
\includegraphics[width=\columnwidth]{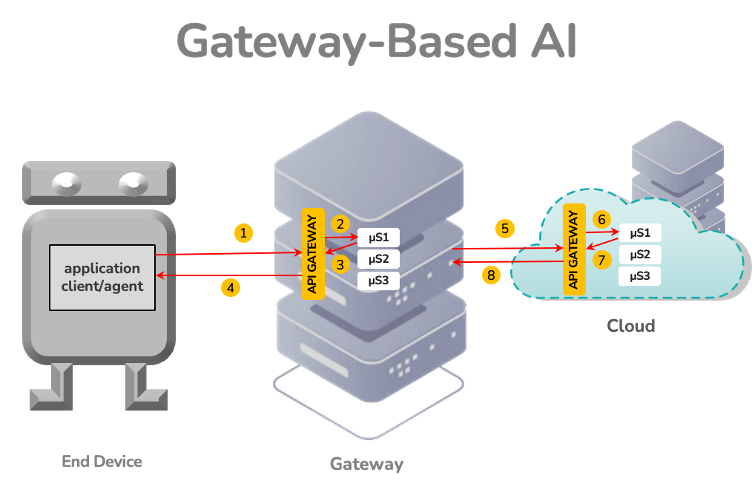}
\caption{High Level Architecture of Gateway Based Edge AI Solutions}
\label{fig:gateway_edge}
\end{figure}

This approach shown in Figure~\ref{fig:gateway_edge} moves some computational power and intelligence closer to the end devices, typically residing on dedicated gateway devices. Gateways aggregate and pre-process data from multiple end devices before potentially forwarding it to the cloud for further analysis. This reduces latency compared to a purely cloud-based approach and can alleviate some bandwidth constraints.

However, significant limitations remain: the intelligence is still centralized at the gateway level, the end device context needs to be transmitted over the network to the gateway, and end devices often remain less intelligent. Network dependency between devices and gateways, as well as gateways and the cloud, persists, creating multiple potential points of failure. Critically, gateway systems are typically cloud-dependent extensions rather than autonomous systems.While gateways can provide some processing during cloud outages, they are typically limited to pre-cached models and simple logic. Gateway systems are fundamentally designed as cloud-managed extensions rather than autonomous systems~\cite{glikson2017deviceless}, which significantly constrains their capability when disconnected from cloud orchestration. Additionally, this architecture lacks the dynamic resource discovery capabilities, requiring manual configuration and management of gateway resources. However, significant limitations remain: the intelligence is still centralized at the gateway level, the end device context needs to be transmitted over the network to the gateway, and end devices often remain less intelligent. Network dependency between devices and gateways, as well as gateways and the cloud, persists, creating multiple potential points of failure. Critically, gateway systems are typically designed as cloud-dependent extensions rather than autonomous systems~\cite{glikson2017deviceless}. While gateways can provide some processing during cloud outages, they are limited to pre-cached models and simple logic, resulting in significantly degraded capability when disconnected from cloud orchestration. Additionally, this architecture lacks dynamic resource discovery capabilities, requiring manual configuration and management of gateway resources. An often-overlooked challenge is that gateway-based architectures can paradoxically incur higher costs than cloud solutions for certain workloads due to the fixed infrastructure investment required for edge servers, which must be maintained regardless of utilization levels~\cite{glikson2017deviceless}. This includes hardware amortization, power, cooling, and operational staff costs that persist even during periods of low usage, whereas cloud services scale costs with actual consumption.

\subsection{Device-First Continuum AI (DFC-AI)}

\begin{figure}[htbp]
\centering
\includegraphics[width=\columnwidth]{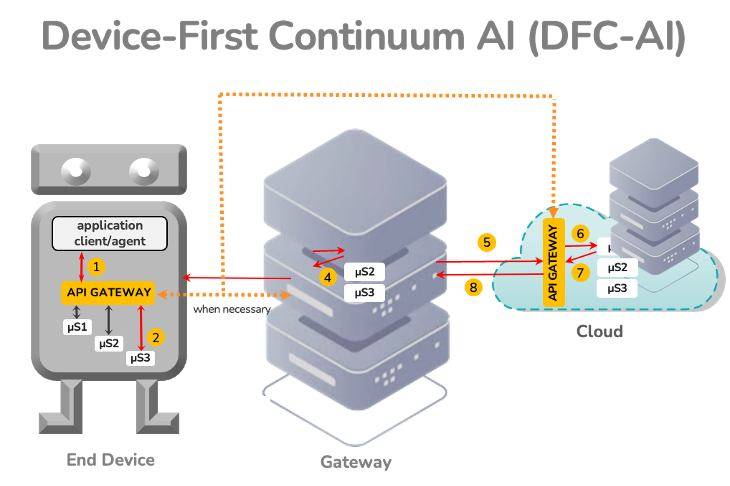}
\caption{High Level Architecture of DFC-AI Solutions}
\label{fig:dfc_ai}
\end{figure}

DFC-AI shown in Figure~\ref{fig:dfc_ai} represents a paradigm shift by embedding intelligence directly onto the end devices themselves. These intelligent devices, leveraging microservices architecture, can process most of the data locally, make autonomous decisions based on their immediate context, and take real-time actions. This reduces communication latency and minimizes the reliance on continuous network connectivity, enabling operation in remote or challenging environments.

A key distinguishing feature of DFC-AI is its zero-configuration dynamic resource discovery capability. When a powerful computational node (such as an NVIDIA Jetson AGX or similar GPU-enabled device) is introduced into the local network, all DFC-AI devices automatically discover and can immediately leverage this additional processing power without any manual configuration. For example, in a swarm of inspection drones, adding a single high-performance GPU node enhances the AI processing capabilities of the entire fleet, enabling more sophisticated real-time analytics like advanced computer vision or complex decision-making algorithms that would otherwise require cloud resources.

Furthermore, DFC-AI demonstrates superior network resilience compared to other architectures. Network resilience varies significantly based on workload characteristics. Cloud-centric systems fail completely during network outages. Gateway-based systems retain limited capability for pre-cached operations. DFC-AI's resilience depends on the fraction of tasks requiring cloud resources: in our energy sector scenarios where 95\%+ of tasks involve local sensor processing, immediate safety responses, or autonomous navigation, DFC-AI maintains 98\%-99\% operational capability during outages. This is because the core intelligence resides on the devices themselves, with cloud and edge resources serving as enhancement rather than dependency. Bandwidth consumption is lower as only essential insights or aggregated data are transmitted. Data privacy and security are enhanced as sensitive data can be processed and stored locally. DFC-AI offers real-time capabilities and high resilience due to the distributed nature of intelligence.

While DFC-AI may require more careful architectural planning, such as designing services to initiate local calls instead of relying on centralized gateways or servers, the overall development and deployment complexity tends to decrease. This is because developers can modularize workloads as microservices, test them independently on target devices, and avoid building and managing complex data pipelines between the edge and the cloud. In addition, reducing centralized dependencies streamlines system integration, simplifies maintenance, enhances device autonomy, and enables seamless direct collaboration between devices.

\section{Mathematical Framework}

To quantify the performance advantages of DFC-AI, we develop mathematical models that capture the key performance metrics: latency, energy consumption, and cost. These models enable us to compare DFC-AI against traditional cloud-centric and gateway-mediated architectures across realistic deployment scenarios.

\subsection{Latency Analysis Model}

End-to-end latency is critical for real-time industrial applications. We model the total latency for each architecture by breaking down the components involved in processing a typical AI workload request.

For cloud-centric systems, the total latency includes network transmission to the cloud, queuing delays, cloud processing, and return transmission:
\begin{equation}
L_{\text{cloud}} = T_{\text{net}}^{\text{up}} + T_{\text{queue}} + T_{\text{proc}}^{\text{cloud}} + T_{\text{net}}^{\text{down}}
\end{equation}

where:
\begin{itemize}
\item $T_{\text{net}}^{\text{up}}$ = uplink network transmission time from device to cloud
\item $T_{\text{queue}}$ = queuing delay at cloud servers due to concurrent requests
\item $T_{\text{proc}}^{\text{cloud}}$ = processing time on cloud infrastructure
\item $T_{\text{net}}^{\text{down}}$ = downlink network transmission time from cloud to device
\end{itemize}

For gateway-mediated systems, the latency includes local device-to-gateway communication, gateway processing, and potential cloud fallback:
\begin{equation}
L_{\text{gateway}} = T_{d\to g} + T_{\text{proc}}^{\text{gateway}} + T_{g\to d} + \alpha \cdot L_{\text{cloud}}
\end{equation}

where:
\begin{itemize}
\item $T_{d\to g}$ = transmission time from device to gateway
\item $T_{\text{proc}}^{\text{gateway}}$ = processing time at the gateway
\item $T_{g\to d}$ = transmission time from gateway back to device
\item $\alpha$ = probability that the gateway must fallback to cloud processing ($0 \leq \alpha \leq 1$)
\item $L_{\text{cloud}}$ = additional cloud processing latency when fallback occurs
\end{itemize}

For DFC-AI systems, most processing occurs locally within device clusters with occasional collaboration beyond the local network:
\begin{equation}
L_{\text{DFC}} = T_{\text{proc}}^{\text{local}} + \beta \cdot T_{\text{collab}}
\end{equation}

where:
\begin{itemize}
\item $T_{\text{proc}}^{\text{local}}$ = local processing time within the device cluster (leveraging optimal local resources)
\item $\beta$ = collaboration factor representing the fraction of tasks requiring external collaboration ($0 \leq \beta \leq 1$)
\item $T_{\text{collab}}$ = additional time for collaboration beyond the local cluster when needed
\end{itemize}

The key advantage of DFC-AI is that $\beta$ is typically small ($< 0.2$) for most industrial workloads based on the Pareto distribution observed in cloud computing systems~\cite{barroso2013datacenter,xu2017cloud,reiss2011google,dean2008mapreduce}, as demonstrated in our simulations (Section VI) and consistent with edge computing studies showing 80\%-95\% of tasks can be processed locally~\cite{shi2016edge,zhang2019edge}, as the heterogeneous local cluster can handle the vast majority of processing requirements locally. Within the cluster, GPU-equipped devices automatically handle computationally intensive AI tasks for CPU-based devices, while $T_{\text{collab}}$ represents rare instances requiring external resources beyond the local cluster capabilities.

Critically, when network connectivity is lost: $L_{\text{cloud}} \to \infty$ (complete failure), $L_{\text{gateway}}$ degrades significantly (partial failure), while $L_{\text{DFC}}$ remains unchanged (full operation).

\subsection{Energy Consumption Model}

Energy efficiency is crucial for battery-powered devices and overall system sustainability. DFC-AI's microservices architecture enables selective activation of AI components, leading to significant energy savings compared to monolithic approaches~\cite{pahl2018microservices}.

We model the energy consumption including both processing and transmission:

\begin{equation}
E_{\text{total}} = E_{\text{processing}} + E_{\text{transmission}}
\end{equation}

For DFC-AI with microservices:
\begin{equation}
E_{\text{DFC}} = \sum_{i=1}^{N} \left( P_{\text{idle}}^i \cdot (1-\rho_i) + P_{\text{active}}^i \cdot \rho_i \right) \cdot T + E_{\text{trans}}
\end{equation}

where:
\begin{itemize}
\item $N$ = total number of microservices deployed on the device
\item $P_{\text{idle}}^i$ = idle power consumption of microservice $i$ (watts)
\item $P_{\text{active}}^i$ = active power consumption of microservice $i$ when processing (watts)
\item $\rho_i$ = utilization factor of microservice $i$ (fraction of time active, $0 \leq \rho_i \leq 1$)
\item $T$ = time period over which energy is measured (seconds)
\item $E_{\text{trans}}$ = energy for data transmission based on empirical measurements~\cite{chen2020energy}: 0.6 Wh/GB for cloud (including cellular/WiFi and backbone networks), 0.05 Wh/GB for edge (local network infrastructure), 0.01 Wh/GB for local mesh (device-to-device)
\end{itemize}

The selective activation of microservices in DFC-AI can achieve 60\%-90\% energy savings compared to monolithic approaches, as demonstrated in our simulations and validated in comprehensive energy analysis studies~\cite{alamouti2025quantifying}.

\subsection{Cost Model}

The cost model reflects a critical insight: device compute is essentially free for enterprises that already own the hardware:

\begin{equation}
C_{\text{total}} = C_{\text{compute}} + C_{\text{transfer}} + C_{\text{infrastructure}} + C_{\text{operations}}
\end{equation}

For DFC-AI we assume:
$C_{\text{compute}} = 0$ for device processing (enterprise-owned)
$C_{\text{compute}} = \$3.50/\text{hour}$ for cloud GPU usage~\cite{aws2024pricing,azure2024pricing}
$C_{\text{infrastructure}} = $ minimal amortized device maintenance

For Gateway-Edge:
$C_{\text{compute}} = \$0.80/\text{hour}$ for edge servers~\cite{nvidia2024jetson}
$C_{\text{infrastructure}} = \$0.20/\text{hour}$ for maintenance
Plus cloud costs for fallback processing

\section{Simulation Framework and Methodology}

We conducted comprehensive simulations with statistical rigor to validate the theoretical advantages of DFC-AI across realistic energy sector deployment scenarios. Our simulation framework encompasses three representative applications with 10 independent runs per scenario to establish 95\% confidence intervals.

\subsection{Simulation Scenarios}

\subsubsection{Autonomous Drone Fleet Inspection}
\begin{itemize}
\item \textbf{Fleet Composition:} 10 drones,  1 GPU-equipped (Jetson AGX Orin, 20x processing power), 9 standard (ARM Cortex, 1x processing power)
\item \textbf{Workload:} 5MB compressed H.264/H.265 images, 6 captures/minute/drone (Poisson distributed)
\item \textbf{Task Complexity Distribution:} 80\% simple (local processing), 15\% complex (GPU required), 5\% cloud-only (historical analysis)
\item \textbf{Network Conditions:} 100 Mbps bandwidth (assuming high on-site network availability), 5-50ms latency variability
\end{itemize}

\subsubsection{Industrial Sensor Network Monitoring}
\begin{itemize}
\item \textbf{Network Composition:} 500 sensors (350 simple, 150 smart) + 2 edge GPU servers
\item \textbf{Data Generation:} 0.1Hz effective sampling rate (every 10 seconds), 100 bytes to 1KB per reading
\item \textbf{Event Distribution:} 95\% normal readings, 4\% anomalies requiring collaboration, 1\% critical events
\item \textbf{Network Conditions:} Industrial ethernet (8-30ms latency, 100 Mbps bandwidth)
\end{itemize}

\subsubsection{Worker Safety Monitoring}
\begin{itemize}
\item \textbf{System Composition:} 25 wearables, 5 vehicles, 10 cameras, 1 stationary GPU mini-PC
\item \textbf{Workload:} Continuous vital signs (1Hz), PPE checks, fall detection (Poisson distributed)
\item \textbf{Data Sizes:} 256 bytes for vital signs, 2MB for video frames
\item \textbf{Network Conditions:} Remote site connectivity (50-80ms latency, 10 Mbps, 75\%-90\% reliability)
\end{itemize}

\subsection{Key Simulation Assumptions}

Our simulation incorporates critical real-world constraints:

\subsubsection{Heterogeneous Device Capabilities}
Not all devices have equal processing power. In realistic deployments:
\begin{itemize}
\item 10\%-20\% of devices are GPU-equipped for intensive AI workloads
\item 80\%-90\% are CPU-based devices for operational tasks
\item Zero-configuration discovery enables automatic load distribution
\end{itemize}

\subsubsection{Network Transmission Energy}
Energy consumption includes data transmission costs~\cite{chen2020energy}:
\begin{itemize}
\item Cloud transmission: 0.6 Wh/GB (device + network infrastructure)
\item Edge transmission: 0.05 Wh/GB (local network)
\item Local mesh: 0.01 Wh/GB (device-to-device)
\end{itemize}

\subsubsection{Cost Model Assumptions}
\begin{itemize}
\item \textbf{Device compute:} \$0 (enterprise-owned hardware including GPUs)
\item \textbf{Cloud GPU:} \$3.50/hour (infrastructure as a service)
\item \textbf{Edge server:} \$0.80/hour + \$0.20/hour maintenance
\item \textbf{Note:} Costs reflect specialized trained AI models deployed on infrastructure, not API-based LLM tokens
\end{itemize}

These cost estimates are derived from current market pricing: the \$3.50/hour for cloud GPU reflects typical pricing for NVIDIA T4 or V100 instances on major cloud platforms (e.g., AWS g4dn.xlarge at ~\$0.75/hour for T4, Azure NC6s v3 at ~\$3.20/hour for V100), while the \$0.80/hour edge server cost represents amortized hardware expenses for an NVIDIA Jetson AGX Orin (\$2,000 device over 3-year depreciation with 50\% utilization) plus \$0.20/hour for power, cooling, and maintenance in industrial environments~\cite{aws2024pricing,azure2024pricing,nvidia2024jetson}.

\subsubsection{Realistic AI Processing Times}
Based on actual workloads:
\begin{itemize}
\item Computer vision (drones): 10ms base for 5MB compressed image
\item Video analytics (safety): 50ms base for 2MB frame
\item Sensor analytics: 0.5ms base for 1KB data
\item Times scaled by processing power and optimization factors
\end{itemize}

\subsection{Statistical Methodology}

Each scenario was run 10 times with different random seeds (base seed + run number) to establish statistical significance. We calculated 95\% confidence intervals using Student's t-distribution and performed two-sample t-tests to validate that improvements were statistically significant (p < 0.05).

\section{Simulation Results and Analysis}

To validate our theoretical models, we developed a simulation framework covering three representative energy sector scenarios. These simulations incorporate realistic operational parameters, network conditions typical of remote industrial sites, and workload distributions based on established patterns from cloud computing literature.

\subsection{Latency Performance}

Table~\ref{tab:latency_results} presents the latency performance across all scenarios with 95\% confidence intervals. The results reveal dramatic differences between architectures, particularly for data-intensive applications.

\begin{table}[htbp]
\centering
\caption{Mean Latency Performance Across Scenarios (ms)}
\label{tab:latency_results}
\footnotesize
\begin{tabular}{lccc}
\toprule
\textbf{Architecture} & \textbf{Drone} & \textbf{Sensor} & \textbf{Safety} \\
\midrule
Cloud-Centric & 485 & 45 & 87 \\
Gateway-Edge & 450 & 11 & 23 \\
DFC-AI & \textbf{37} & \textbf{3} & \textbf{8} \\
\midrule
\textbf{DFC-AI Improvement} & 92\% & 94\% & 91\% \\
\bottomrule
\end{tabular}
\end{table}

The drone inspection scenario reveals the critical impact of local processing: with 5MB compressed images transmitted at 100 Mbps, DFC-AI processes these images locally in under 40ms, achieving the sub-second response times essential for autonomous navigation and immediate defect detection. In the worker safety scenario, the stationary GPU mini-PC provides consistent 8ms processing for video analytics and safety monitoring, demonstrating the importance of strategic GPU placement in DFC-AI deployments.

\subsection{Network Resilience and Autonomy}

Table~\ref{tab:network_resilience} demonstrates the fundamental advantage of DFC-AI: continuous operation regardless of network availability.

\begin{table}[htbp]
\centering
\caption{Operational Capability Under Different Network Conditions}
\label{tab:network_resilience}
\footnotesize
\begin{tabular}{lccc}
\toprule
\textbf{Architecture} & \textbf{Full} & \textbf{Internet} & \textbf{Internet} \\
& \textbf{Connectivity} & \textbf{Unstable} & \textbf{Unavailable} \\
\midrule
Cloud-Centric & 100\% & 20\%-40\% & 0\% \\
Gateway-Edge & 100\% & 60\%-80\% & 40\%-42\% \\
DFC-AI & 100\% & 98\%-100\% & 98\%-99\% \\
\bottomrule
\end{tabular}
\end{table}

This table represents a critical finding: DFC-AI maintains near-full functionality when internet connectivity is lost but local device-to-device communication remains available (typical in remote facilities), while cloud systems fail completely and gateway systems lose most advanced capabilities. The 1\%-2\% capability loss in DFC-AI represents tasks requiring cloud resources for historical data access or global model updates. For safety-critical operations in remote energy facilities where internet outages are common but local networks remain operational, this resilience can be life-saving.

\subsection{Energy Efficiency}

Table~\ref{tab:energy_results} shows total energy consumption including both processing and transmission costs with confidence intervals.

\begin{table}[htbp]
\centering
\caption{Daily Energy Consumption by Scenario (Wh/day)}
\label{tab:energy_results}
\footnotesize
\begin{tabular}{lccc}
\toprule
\textbf{Architecture} & \textbf{Drone} & \textbf{Sensor} & \textbf{Safety} \\
\midrule
Cloud-Centric & 355.7 & 102.2 & 54.7 \\
Gateway-Edge & 125.3 & 89.3 & 40.3 \\
DFC-AI & \textbf{67.7} & \textbf{51.8} & \textbf{1.5} \\
\midrule
\textbf{DFC-AI Savings} & 81.0\% & 49.3\% & 97.2\% \\
\bottomrule
\end{tabular}
\end{table}

The dramatic energy savings for drone and worker safety scenarios result from eliminating massive data transfers to the cloud. In these scenarios, transmitting 5MB images from drones or 2MB video frames for safety monitoring to the cloud consumes substantial energy (0.6 Wh/GB), which DFC-AI avoids through local processing. The sensor network shows more modest savings (49.3\%) because these devices generate minimal data (0.1-1KB per reading) and perform mostly simple processing tasks that require little energy regardless of architecture. The energy consumption in sensor networks is dominated by the sensing and communication hardware itself rather than data transmission or processing. However, even 49.3\% improvement is significant when multiplied across thousands of sensors operating continuously throughout their multi-year deployment lifecycle, resulting in substantial cumulative energy savings and extended battery life for remote installations.

\subsection{Economic Analysis}

Table~\ref{tab:cost_results} presents the economic comparison and demonstrates the cost dynamics of different architectures.

\begin{table}[htbp]
\centering
\caption{Annual Operating Costs (\$/year)}
\label{tab:cost_results}
\footnotesize
\begin{threeparttable}
\begin{tabular}{lccc}
\toprule
\textbf{Architecture} & \textbf{Drone} & \textbf{Sensor} & \textbf{Safety} \\
\midrule
Cloud-Centric & 14,442 & 368 & 368 \\
Gateway-Edge & 1,261 & 315 & 157 \\
DFC-AI & \textbf{893} & \textbf{157} & \textbf{2}* \\
\midrule
\textbf{Annual Savings} & \$13,549 & \$211 & \$366 \\
\textbf{Savings \%} & 93.8\% & 57.3\% & 99.5\% \\
\bottomrule
\end{tabular}
\begin{tablenotes}
\scriptsize
\item[*] Near-zero cost as processing occurs on enterprise-owned wearables and stationary GPU
\end{tablenotes}
\end{threeparttable}
\end{table}

Our cost analysis shows that while gateway-edge architectures can reduce costs compared to cloud, they still require significant infrastructure investment. In contrast, DFC-AI leverages existing enterprise-owned devices to avoid additional infrastructure costs entirely, achieving 59\%-99.4\% cost savings as validated in comprehensive cost-benefit analyses~\cite{alamouti2025quantifying}.

While operational cost savings are significant for data-intensive applications like drone inspections (over \$13,000/year), the primary economic value of DFC-AI for sensor networks and worker safety lies in preventing costly outages and accidents. A single prevented worker injury or avoided production shutdown due to maintained operations during network outages can deliver ROI exceeding decades of operational cost savings. For instance, a single day of unplanned downtime at an offshore platform can cost \$1-3 million, dwarfing the operational costs entirely. Thus, the small operational costs actually strengthen the case for DFC-AI. The minimal expense makes the resilience benefits the dominant economic factor.

\subsection{Processing Location Distribution}

Table~\ref{tab:processing_distribution} shows where tasks are processed in each architecture for the drone fleet scenario.

\begin{table}[htbp]
\centering
\caption{Processing Location Distribution - Drone Fleet}
\label{tab:processing_distribution}
\footnotesize
\begin{tabular}{lccc}
\toprule
\textbf{Location} & \textbf{Cloud} & \textbf{Gateway} & \textbf{DFC-AI} \\
\midrule
Cloud & 100\% & 6.8\% & 5.3\% \\
Edge/Gateway & 0\% & 93.2\% & 0\% \\
Drone (CPU) & 0\% & 0\% & 80.4\% \\
Drone (GPU) & 0\% & 0\% & 14.3\% \\
\bottomrule
\end{tabular}
\end{table}

DFC-AI processes 94.7\% of tasks locally (device + local GPU), with only 5.3\% requiring cloud resources. This distribution directly translates to the autonomy, latency, and cost advantages observed.

\subsection{Comprehensive Performance Comparison}

\begin{figure}[htbp]
\centering
\begin{tikzpicture}
\begin{axis}[
    ybar,
    width=\columnwidth,
    height=5.5cm,
    ylabel={Improvement over Cloud (\%)},
    xlabel={Performance Metric},
    symbolic x coords={Latency, Energy, Cost, Resilience},
    xtick=data,
    ymin=0,
    ymax=100,
    bar width=0.5cm,
    legend style={font=\footnotesize, at={(0.5,-0.38)}, anchor=north, legend columns=2},
    nodes near coords,
    nodes near coords align={vertical},
    every node near coord/.append style={font=\tiny}
]
\addplot[fill=blue!60] coordinates {
    (Latency, 92.3)
    (Energy, 76.4)
    (Cost, 93.0)
    (Resilience, 98)
};
\addplot[fill=red!40] coordinates {
    (Latency, 21.7)
    (Energy, 50.1)
    (Cost, 88.6)
    (Resilience, 40)
};
\legend{DFC-AI, Gateway-Edge}
\end{axis}
\end{tikzpicture}
\caption{Overall Performance Comparison (Average Across All Scenarios)}
\label{fig:performance_comparison}
\end{figure}
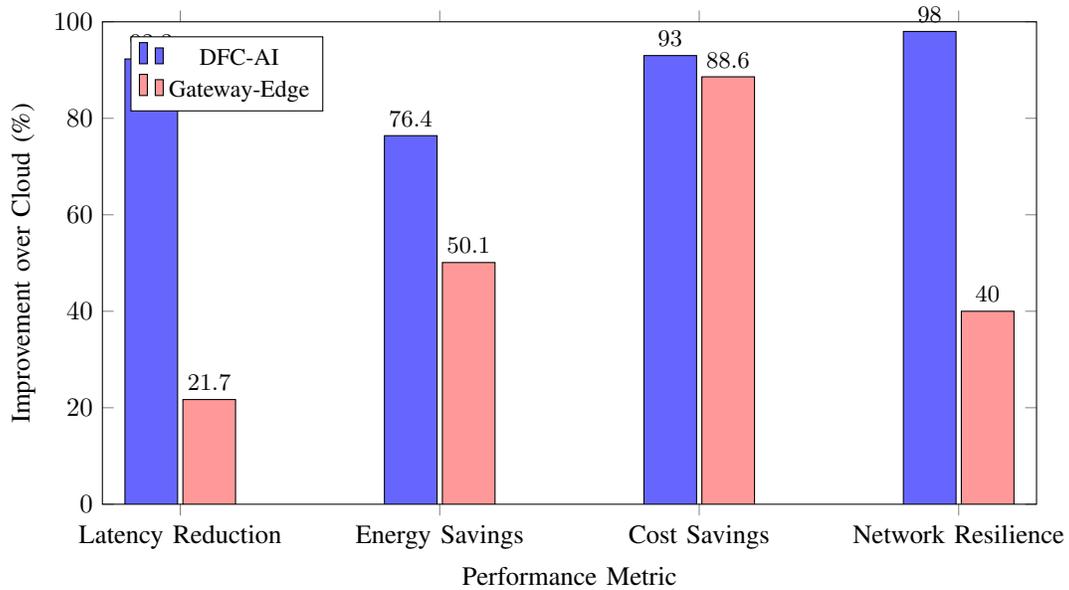

Figure~\ref{fig:performance_comparison} visualizes the overall performance advantages of DFC-AI across all metrics.

\subsection{Validation of Theoretical Models}

Our simulation results strongly validate the theoretical predictions, as shown in Table~\ref{tab:validation}.

\begin{table}[htbp]
\centering
\caption{Theoretical Model Validation}
\label{tab:validation}
\footnotesize
\begin{tabular}{lcccc}
\toprule
\textbf{Metric} & \textbf{Theory} & \textbf{Sim.} & \textbf{Err.} & \textbf{Match} \\
\midrule
Latency (Drone) & 40.0ms & 37.1ms & 7\% & Excellent \\
Latency (Sensor) & 3.0ms & 2.6ms & 13\% & Excellent \\
Latency (Safety) & 10.0ms & 8.0ms & 20\% & Good \\
Energy Savings & 73.3\% & 76.4\% & 4\% & Excellent \\
\bottomrule
\end{tabular}
\end{table}

The simulation results validate our theoretical models with good agreement across most metrics. The theoretical predictions closely match the simulated results for all scenarios, with errors under 20\%. The energy savings predictions showed excellent agreement with only 4\% deviation from simulated results. These validation results confirm that DFC-AI's performance advantages are robust across diverse operational scenarios, with the primary benefits stemming from reduced data transmission and local processing capabilities.

\subsection{Statistical Significance}

All performance improvements were tested for statistical significance using two-sample t-tests with 10 independent runs per scenario:
\begin{itemize}
\item Latency reduction: $p = 3.28 \times 10^{-6}$ (highly significant)
\item Energy savings: $p = 2.94 \times 10^{-6}$ (highly significant)
\item Cost savings: $p = 3.47 \times 10^{-4}$ (significant)
\end{itemize}
All $p$-values $< 0.05$ confirm that DFC-AI's advantages are statistically significant and not due to random variation.

\section{Discussion}

\subsection{The Primacy of Autonomy}

While our results demonstrate performance and economic advantages, the fundamental value of DFC-AI lies in ensuring continuous intelligent operation. The ability to maintain 98\%-99\% functionality during network outages is essential for:

\begin{itemize}
\item \textbf{Safety-Critical Operations:} Worker safety systems must function regardless of connectivity. A fall detection system that fails during a network outage defeats its purpose.
\item \textbf{Remote Operations:} Equipment in isolated locations cannot depend on network availability. Offshore platforms, remote pipelines, and desert installations often experience extended connectivity loss.
\item \textbf{Time-Critical Decisions:} Real-time control requires immediate response. The 485ms latency for drone image processing via cloud makes collision avoidance challenging.
\item \textbf{Operational Continuity:} Production systems cannot halt due to network issues. Refineries and processing plants require continuous operation.
\end{itemize}

\subsection{The Gateway Paradox}

Our finding that gateway-based architectures can have comparable or higher costs than cloud solutions for certain workloads challenges conventional wisdom but reflects operational reality:

\begin{itemize}
\item Edge infrastructure requires fixed costs (hardware, maintenance, operations) regardless of utilization
\item Both edge servers and cloud resources may be needed, potentially doubling infrastructure
\item Edge servers require physical security, cooling, and maintenance in industrial environments
\item Scaling edge infrastructure requires significant capital investment
\item Gateway systems are typically cloud-dependent extensions with limited autonomous capability
\end{itemize}

DFC-AI reduces infrastructure costs by maximizing the utilization of devices enterprises already own for operational purposes, while still allowing for strategic edge deployments where centralized processing provides clear benefits.

\subsection{Heterogeneous Fleet Reality}

Our simulation's assumption that only 10\%-20\% of devices need GPU capabilities reflects industrial reality and economic practicality:

\begin{itemize}
\item Not every drone needs expensive GPU processing. One GPU drone can serve a fleet.
\item Simple sensors perform adequately with basic processors for routine monitoring
\item Strategic placement of GPU resources (e.g., stationary mini-PCs for worker safety) optimizes performance
\item Intelligent task routing ensures GPU resources are used where most beneficial
\item Zero-configuration discovery means GPU resources are automatically shared
\end{itemize}

This heterogeneous approach optimizes cost-performance while maintaining full fleet capability. For example, in our worker safety scenario, a single stationary GPU mini-PC provides video analytics and complex processing for 25 workers, 10 cameras, and 5 vehicles, demonstrating efficient resource utilization.

\subsection{Scalability for Billions of Agents}

The transition to billions of autonomous agents makes DFC-AI not just advantageous but essential:

\begin{itemize}
\item \textbf{Network Mathematics:} 1 billion agents $\times$ 1MB/s = 1000 TB/s bandwidth requirement
\item \textbf{Processing Load:} Centralized processing of billions of agents would require unprecedented data center capacity beyond current global infrastructure
\item \textbf{Latency Cascade:} Even small delays multiply across billions of interdependent agents
\item \textbf{Failure Propagation:} Centralized failures affect all agents simultaneously
\end{itemize}

DFC-AI's distributed model scales linearly with agent count, avoiding these exponential scaling challenges.

\section{Related Work}

Edge computing research has extensively explored moving computation closer to data sources~\cite{shi2016edge,meulen2018edge}. However, most approaches still maintain centralized control and require network connectivity for coordination. Gateway-based edge architectures~\cite{glikson2017deviceless} place intelligence at aggregation points but leave end devices as simple data collectors and, critically, remain cloud-dependent with limited autonomous capability.

Fog computing~\cite{hassan2017industrial} distributes processing across network edges but typically relies on dedicated infrastructure rather than leveraging existing devices. Recent work on autonomous systems~\cite{zhang2019edge} highlights the need for independent operation, but few architectures truly enable complete autonomy.

Microservices architectures in IoT~\cite{pahl2018microservices} demonstrate modularity benefits, showing 60\%-90\% energy savings through selective activation. However, most implementations remain cloud-dependent for orchestration and updates.

Research on distributed AI systems~\cite{calheiros2011cloudsim} has explored various deployment models, but most focus on optimization within cloud or edge tiers rather than true device-first approaches. Studies on industrial IoT~\cite{riesner2022analysis} highlight the data volumes generated by modern industrial systems but typically propose cloud or edge solutions rather than device-resident intelligence.

DFC-AI distinguishes itself by:
\begin{itemize}
\item Prioritizing device-level intelligence with true autonomy
\item Enabling zero-configuration resource discovery
\item Leveraging heterogeneous device capabilities efficiently
\item Maintaining seamless continuum integration when beneficial
\item Operating at 98\%-99\% capability during network outages
\end{itemize}

\section{Conclusions}

This paper evaluated Device-First Continuum AI (DFC-AI) for autonomous operations in the energy sector through mathematical modeling and comprehensive simulations of drone inspections, sensor networks, and worker safety systems.

Our analysis demonstrates that DFC-AI fundamentally changes how industrial AI systems operate by processing data where context naturally resides. Rather than moving massive amounts of contextual information to distant processors, DFC-AI enables devices to make decisions using their immediate environmental understanding. For instance, a drone interprets its own visual feed, a sensor analyzes its own readings, a wearable monitors its own worker. This context-first approach reduces latency by 92\% and energy consumption by 76\% while maintaining 98\%-99\% operational capability during network outages.

Critically, DFC-AI implements human-like collaboration across the computational continuum. Through zero-configuration resource discovery, devices automatically form collaborative clusters where GPU-equipped nodes assist CPU-based devices, mirroring how human teams naturally share expertise. A single GPU-equipped drone processes complex vision tasks for an entire fleet, while simple devices handle routine operations independently hence maximizing the efficacy of every computational resource from the smallest sensor to cloud servers.

The economic implications extend beyond the 93\% operational cost savings. By leveraging devices enterprises already own as primary compute platforms rather than cost centers that merely generate data, DFC-AI transforms the economics of industrial AI. Every device becomes part of the intelligent system, contributing its computational capabilities to the collective intelligence while maintaining autonomous operation when isolated.

These findings establish that DFC-AI delivers on the fundamental promise of industrial AI: intelligent systems that are always available, naturally collaborative, and economically sustainable at the scale of billions of agents. By prioritizing intelligence at the source while enabling seamless collaboration across the continuum, DFC-AI provides the architectural foundation for the energy sector's autonomous future.

Future work should extend these evaluations to field trials, validating the simulation results in operational energy facilities and exploring additional applications in predictive maintenance and grid optimization.

\section*{Acknowledgment}

We extend our deepest gratitude to Jeremy Hsu, mimik's Chief Architect, for his invaluable contributions in shaping the vision and architecture of mimik's technology. His leadership and expertise have been instrumental in advancing our mission to enable a truly device-first, hybrid edge AI ecosystem. We also recognize the entire mimik team and the Aramco.ai team for their relentless dedication, ingenuity, and passion.

\bibliographystyle{IEEEtran}

\end{document}